\documentstyle[preprint,aps]{revtex} 

\begin{document}

\draft

\tighten


\title{On Quantum Corrections to Chern-Simons Spinor Electrodynamics}
\author{M. Chaichian${}^{a,b}$ W.F. Chen${}^b$\thanks{ICSC-World Laboratory,
Switzerland} and V.Ya. Fainberg${}^c$}

\address{${}^a$ High Energy Physics Division,
Department of Physics\\ 
${}^b$ Helsinki Institute of Physics\\
P.O. Box 9 (Siltavourenpenger 20 C), FIN-00014 University of
Helsinki, Finland\\
${}^c$ P.N. Lebedev Institute of Physics, Moscow, Russia}


\maketitle

\begin{abstract}
We make a detailed investigation on the quantum corrections to 
Chern-Simons spinor electrodynamics.
Starting from Chern-Simons spinor quantum electrodynamics 
with the Maxwell term $-1/(4\gamma){\int}d^3x F_{\mu\nu}F^{\mu\nu}$
and by calculating the vacuum polarization tensor, electron self-energy
and on-shell vertex, we explicitly show that the Ward identity is 
satisfied and hence  verify that
the physical quantities are independent of the procedure of 
taking ${\gamma}{\to}{\infty}$
at tree and one-loop levels. 
In particular, we find the three-dimensional
analogue of the  Schwinger anomalous magnetic moment term of the electron
produced from the quantum corrections.
\end{abstract}

\eject

\section{Introduction}

There is a relatively long history for (Abelian or Non-Abelian)
Chern-Simons (CS) theory and its relevant theories to become popular
in physics. At early stage they appeared as the 
high-temperature limit of four dimensional field models, where
Maxwell-Chern-Simons theory can be regarded as an effective theory
of QCD and the electroweak model 
\cite{gpy}. Further its more striking aspect 
had been found: in three-dimensional space-time 
the CS term can provide a topological
mass for the gauge field in a gauge invariant way as an alternative
to Higgs mechanism\cite{sc,djt}. In recent years the revival to the study
of CS theory, on one hand, is due to Witten's work\cite{wit}
in which a connection between
CS theory and 2-dimensional conformal invariant field theory 
was found; 
on the other hand, owing to the non-invariance 
of CS term under $P$ and $T$ transformations and especially
its topological character, it can be used to describe the dynamics of
anyon particles so that it has been favoured by physicists to solve some 
problems in condensed
matter theory such as the fractional quantum Hall effect and the
high temperature superconductivity\cite{col}. It is also proved that  
CS term coupled to scalar matter is useful in
the field-theoretic formulation  of the Aharonov-Bohm effect\cite{fa}
and the three-dimensional analogue of Coleman-Weinberg mechanism is
explored up to two-loop\cite{tth}.

In this letter we shall present a detailed investigation on the one-loop
quantum correction of one-loop CS spinor electrodynamics.
We start from the action with Maxwell term
\begin{eqnarray}
S&=&\frac{\mu}{2} \int d^3x \,\epsilon^{\mu\nu\lambda} A_\mu\partial_\nu
A_\lambda - \frac{1}{4\gamma}\int d^3x\, F_{\mu\nu}F^{\mu\nu}\nonumber \\[2mm]
&+&\int d^3x\,\left[\bar{\psi}(i\hat{\partial}+e\hat{A}-m)\psi\right]
-\frac{1}{2\alpha}\int d^3x(\partial_\mu A^\mu)^2,
\label{eqac}
\end{eqnarray}
where (and in what follows) $\hat{A}{\equiv}{\gamma}^{\alpha}A_{\alpha}$, 
 $\mu$ is the statistical parameter and we choose the Lorentz
gauge condition ${\partial}_{\mu}A^{\mu}=0$. The notation is
the same as that in Ref.{\cite{djt}},
\begin{eqnarray}
{\gamma}_{\mu}=i{\sigma}_{\mu}, ~~{\gamma}_{\mu}{\gamma}_{\nu}=
g_{\mu\nu}-i{\epsilon}_{\mu\nu\rho}{\gamma}^{\rho},~~
g_{\mu\nu}=\mbox{diag}(1,-1,-1).
\label{eqno}
\end{eqnarray}
It should be stressed
that the introduction of Maxwell term plays a two-fold role: on one hand,
it provides a mathematically correct path-integral
quantization of the CS theory in Euclidean region, since the pure CS term
contains the non-positive definite first order differential operator;
on the other hand, as a higher order derivative term, it provides
a gauge invariant regularization. However, this regularization is not
enough to make one-loop amplitude finite, another regularization must be
implemented. Here we shall adopt dimensional regularization.

The model (\ref{eqac})
 has been studied by many authors$\cite{kao}$, especially 
the case where CS term is absent (i.e. pure QED${}_3$). However
they mainly consider the dynamical mass generation,  the chiral
symmetry and parity breaking by quantum corrections. A complete
investigation on its quantum correction still lacks such as the
explicit verification of Ward identity and whether there exists the
three-dimensional analogue of Schwinger's anomalous magnetic moment term,
all of which depend on an explicit analytical calculation of the
vertex correction. To our knowledge, up to now there appears
no analytic result on this part. We are further motivated by the result of
pure non-Abelian CS theory, where a different order
in taking ${\gamma}{\to}{\infty}$ 
can result in a different finite renormalization of
the statistical parameter\cite{aso}. It is desirable 
to see whether this case happens
in CS spinor electrodynamics too. As we know, in quantum
electrodynamics, the Ward identity means
\begin{eqnarray}
Z_1=Z_2,
\end{eqnarray}
where $Z_2$ and $Z_1$ are the electron wave
function renormalization constant and vertex renormalization
constant respectively. So if the Ward identity is satisfied, the renormalization
of coupling constant is only relevant to the gauge field 
wave function renormalization
constant $Z_3$:
\begin{eqnarray}
e_R=\sqrt{Z_3}Z_1Z_2^{-1}e=\sqrt{Z_3}e.
\end{eqnarray}
Since $Z_3$ is independent of the introduction of Maxwell term,
so the Ward identity means that the physical quantities have nothing to do with
the order of taking ${\gamma}{\to}{\infty}$. In particular it is very
interesting to see whether there exists an anomalous magnetic moment
term, since it can produce a new interaction between anyons
that will lead to unusual planar dynamics$\cite{ck,ks}$. 
This may be helpful to understand the mechanisms of fractional quantum Hall effects and high 
temperature superconductivity.
      
The Feynman rules are listed as follows
\begin{itemize}
\item gauge field propagator
\begin{eqnarray}
\tilde{D}^{(0)}_{\mu\nu}(p)=-i\frac{\gamma}{p^2-{\mu}^2{\gamma}^2}\left[
i{\mu}{\gamma}{\epsilon}_{\mu\nu\rho}\frac{p^{\rho}}{p^2}+
g_{\mu\nu}-\frac{p_{\mu}p_{\nu}}{p^2}\right],
\label{eq5}
\end{eqnarray}
where we choose Landau gauge (${\alpha}=0$) to avoid infrared
singularity\cite{djt,rao}. In the limit of ${\gamma}{\to}{\infty}$, we have
\begin{eqnarray}
D^{(0)}_{\mu\nu}(p)
=-\frac{1}{\mu}{\epsilon}_{\mu\nu\rho}\frac{p^{\rho}}{p^2}.
\label{eq6}
\end{eqnarray}
\item electron propagator
\begin{eqnarray}
S^{(0)}(p)=i\frac{\hat{p}+m}{p^2-m^2}.
\end{eqnarray}
\item the vertex
\begin{eqnarray}
-ie{\gamma}_{\mu}(2\pi)^3 {\delta}^{(3)}(p+q+r).
\end{eqnarray}
\end{itemize} 

In Sect. II, starting from the classical action (\ref{eqac}), we 
calculate the vacuum polarization tensor and electron
self-energy correction and define the finite renormalization
constants relevant to them. In Sect. III, we compare
our results obtained in dimensional regularization 
with those obtained in Pauli-Villars regularization
and expressed in spectral representation and find
the results are identical, this shows the gauge invariant 
regularization scheme independence.  Sect. IV is devoted to a detail
calculation of mass-shell vertex correction. 
It is explicitly shown that the Ward identity is 
satisfied on mass-shell. Especially, we find the three dimensional
analogue of anomalous magnetic moment term. In Sect. V we turn
to pure CS spinor electrodynamics (i.e. taking
${\gamma}{\rightarrow}{\infty}$ at tree level) and we verify that
the Ward identity is still satisfied, which shows that the physical
quantities are independent of the order of taking large-$\gamma$ limit.
Sect. VI contains the conclusions and some 
discussions on higher order results.

\section{One-loop Vacuum Polarization and Self-energy}

\subsection{Polarization tensor}

The polarization tensor gets contribution from the electron loop and
its amplitude is
\begin{eqnarray}
i{\Pi}_{\mu\nu}(p)
=-2e^2{\int}\frac{d^nq}{(2\pi)^n}
\frac{-im{\epsilon}_{\mu\nu\rho}p^{\rho}+2q_{\mu}q_{\nu}+q_{\mu}p_{\nu}+
q_{\nu}p_{\mu}+[m^2-q{\cdot}(q+p)]g_{\mu\nu}}{(q^2-m^2)[(q+p)^2-m^2]}.
\end{eqnarray}
The standard calculation gives 
\begin{eqnarray}
{\Pi}_{\mu\nu}(p)&=&i{\epsilon}_{\mu\nu\rho}p^{\rho}{\Pi}_{\rm o}(p^2)+
(p^2g_{\mu\nu}-{p_{\mu}p_{\nu}}){\Pi}_{\rm e}(p^2)\nonumber\\[2mm]
&=&\frac{e^2}{4{\pi}}\left\{i{\epsilon}_{\mu\nu\rho}p^{\rho}
\frac{m}{p}\ln\frac{1+p/(2m)}{1-p/(2m)}\right.\nonumber\\[2mm]
&-&\left.
(p^2g_{\mu\nu}-p_{\mu}p_{\nu})\left[
-\frac{m}{p^2}+\left(\frac{1}{4p}
+\frac{m^2}{p^3}\right)\ln\frac{1+p/(2m)}{1-p/(2m)}\right]
\right\}. 
\label{eq10}
\end{eqnarray}

\subsection{Electron Self-energy}

The Feynman integral for electron self-energy is read as follows
\begin{eqnarray}
-i\tilde{\Sigma}(p,m,{\gamma})
=-e^2{\gamma}{\int}\frac{d^nq}{(2\pi)^n}
\frac{-\hat{q}[m{\mu}{\gamma}+q{\cdot}(q+p)]+q^2m+{\mu}{\gamma}q{\cdot}(q+p)}
{[(q+p)^2-m^2]q^2(q^2-{\mu}^2{\gamma}^2)}.
\label{eqse}
\end{eqnarray}
Using the identities
\begin{eqnarray}
\frac{1}{q^2(q^2-{\mu}^2{\gamma}^2)}&=&\frac{1}{{\mu}^2{\gamma}^2}
(\frac{1}{q^2-{\mu}^2{\gamma}^2}-\frac{1}{q^2}),\nonumber\\[2mm]
2q{\cdot}p&=&[(q+p)^2-m^2]-q^2-(p^2-m^2)\nonumber\\[2mm]
&=&[(q+p)^2-m^2]-(q^2-{\mu}^2{\gamma}^2)-
(p^2-m^2+{\mu^2}{\gamma}^2),
\end{eqnarray}
Eq.(\ref{eqse}) can be written as
\begin{eqnarray}
-i\tilde{\Sigma}(p,m,{\gamma})&=&-2e^2{\gamma}{\int}
\frac{d^nq}{(2\pi)^n}\left\{\left(m+\frac{{\mu}{\gamma}}{2}-
\frac{p^2-m^2}{2{\mu}{\gamma}}\right)
\frac{1}{(q^2-{\mu}^2{\gamma}^2)[(q+p)^2-m^2]}\right.\nonumber\\[2mm]
&+&\frac{1}{2{\mu}{\gamma}}\frac{1}{q^2-{\mu}^2{\gamma}^2}+
\frac{p^2-m^2}{2{\mu}{\gamma}}\frac{1}{q^2[(q+p)^2-m^2]}\nonumber\\[2mm]
&-&\left(\frac{m}{\mu\gamma}+\frac{1}{2}-\frac{p^2-m^2}{2{\mu}^2{\gamma}^2}
\right)
\frac{\hat{q}}{(q^2-{\mu}^2{\gamma}^2)[(q+p)^2-m^2]}
\nonumber\\[2mm]
&-&\left.\left(\frac{p^2-m^2}{2\mu^2{\gamma}^2}-\frac{m}{\mu\gamma}\right)
\frac{\hat{q}}{q^2[(q+p)^2-m^2]}\right\}.
\end{eqnarray}
After the integration and the limit of ${\gamma}{\to}{\infty}$,
we have
\begin{eqnarray}
{\Sigma}(p)
&=&\lim_{{\gamma}{\rightarrow}\infty}\tilde{\Sigma}(p, m,
\gamma)=\frac{e^2}{4\pi}\left\{2{\gamma}
+\frac{m}{\mu}+\frac{p^2-m^2}{{\mu}p}\ln\frac{1+p/m}{1-p/m}\right.
\nonumber\\[2mm]
&-&\left.
\frac{\hat{p}}{\mu}\left[\frac{m^2}{p^2}+\frac{m}{p}(1-\frac{m^2}{p^2})
\ln\frac{1+p/m}{1-p/m}-\frac{2}{3}\right]\right\}. 
\label{eq14}
\end{eqnarray}

\subsection{Finite Renormalization}

Now we discuss the finite renormalization of one-loop
two point functions.
From 
\begin{eqnarray}
D^{(1)\,-1}_{\mu\nu}(p)=D^{(0)\,-1}_{\mu\nu}(p)-i{\Pi}_{\mu\nu}(p),
\end{eqnarray}
we can get the one-loop gauge field propagator 
\begin{eqnarray}
D^{(1)}_{\mu\nu}(p)&=& -i(g_{\mu\nu}-\frac{p_{\mu}p_{\nu}}{p^2})
\frac{{\Pi}_{\rm e}(p^2)}{{\mu}^2[1-{\Pi}_{\rm o}(p^2)]^2
-p^2{\Pi}^2_{\rm e}(p^2)}\nonumber\\[2mm]
&-&{\epsilon}_{\mu\nu\rho}\frac{p^{\rho}}{p^2}\frac{\mu [1-\Pi_{\rm o}(p^2)]}
{{\mu}^2[1-{\Pi}_{\rm o}(p^2)]^2-p^2{\Pi}_{\rm e}^2(p^2)}.
\end{eqnarray}
The renormalized propagator should be the following form
\begin{eqnarray}
D^{(1)}_{\mu\nu}(p)=-i(g_{\mu\nu}-\frac{p_{\mu}p_{\nu}}{p^2}){\Pi}_1(p^2)
-\epsilon_{\mu\nu\rho}\frac{p^{\rho}}{p^2}
\left[\frac{Z_3}{\mu_{\rm ph}}+\Pi_2(p^2)\right].
\end{eqnarray}
 Choosing the renormalization point  $p^2=0$, we get 
\begin{eqnarray}
Z_3&=&1,\nonumber\\[2mm] 
{\mu}_{\rm ph}&=&{\mu}(1-\frac{e^2}{4\pi}).
\end{eqnarray}
Especially, one can see ( up to the order $e^2$)
\begin{eqnarray}
\Pi_1(0)=-\frac{e^2}{4\pi}\frac{1}{3m}{\neq}0,
\end{eqnarray}
which means the quantum correction generates the parity-even part
of the gauge field propagator.

As for the finite renormalization of electron self-energy, 
it is defined by the usual mass-shell renormalization condition
\begin{eqnarray}
{\Sigma}_R({p})|_{\hat{p}=m_{\rm ph}}=0, 
~~\frac{\partial}{{\partial}\hat{p}}{\Sigma}_R(p)|_{\hat{p}=m_{\rm ph}}=0.
\end{eqnarray}
Thus the self-energy can be written as the expansion around
$\hat{p}=m_{\rm ph}$,
\begin{eqnarray}
{\Sigma}(p)={\delta}m-(Z_2^{-1}-1)(\hat{p}-m_{\rm ph})+Z_2^{-1}{\Sigma}_R(p)
\end{eqnarray}
and the one-loop electron propagator is 
\begin{eqnarray} 
S^{(1)}(p)=i\frac{Z_2}{\hat{p}-m_{\rm ph}-{\Sigma}_R(p)}
=i\left[\frac{Z_2}{\hat{p}-m_{\rm ph}}+\tilde{\Sigma}_R(p)\right].
\label{eq3}
\end{eqnarray}
From the one-loop correction (\ref{eq14}),
the physical mass, electron wave function
renormalization constant and the radiative correction are
(up to the order $e^2$) given by
\begin{eqnarray}
m_{\rm ph}&=&m-{\delta}m=m-\frac{e^2}{2\pi}({\gamma}+\frac{m}{3\mu}),
\nonumber\\[2mm]
Z_2&=&1+\frac{e^2}{4\pi}\frac{5}{3\mu},\nonumber\\[2mm]
{\Sigma}_R(p)&=& \frac{e^2}{4\pi}\left\{
\frac{2m_{\rm ph}}{\mu}+\frac{p^2-m_{\rm ph}^2}{{\mu}p}
\ln\frac{1+p/m_{\rm ph}}{1-p/m_{\rm ph}}\right.\nonumber\\[2mm]
&-&\left.\frac{\hat{p}}{\mu}\left[1+\frac{m^2_{\rm ph}}{p^2}
+\frac{m_{\rm ph}}{p}(1-\frac{m^2_{\rm ph}}{p^2})
\ln\frac{1+p/m_{\rm ph}}{1-p/m_{\rm ph}}\right]\right\},\nonumber\\[2mm]
\tilde{\Sigma}_R(p)&=&\frac{e^2}{4\pi\mu}\frac{\hat{p}}{p^2}\left(1-
\frac{\hat{p}+m_{\rm ph}}{2p}\ln\frac{1+p/m_{\rm ph}}{1-p/m_{\rm ph}}\right).
\label{eq25}
\end{eqnarray}

\section{Comparison with the Results in Spectral Representation}

In Ref.$\cite{djt}$, the one-loop two point functions of CS 
spinor electrodynamics had been  presented in terms of the 
spectral representation. Regarding the Maxwell term as a higher covariant 
derivative term, we can consider the results in Ref.$\cite{djt}$ obtained
by Pauli-Villars
regularization. If the large topological mass limit is taken,  their results
should be consistent with ours since both regularization schemes are
gauge invariant. The aim of this section is to show it explicitly.

\subsection{Polarization Tensor}


We start from Eqs.(2.61)--(2.64b) of Ref.$\cite{djt}$.
After the renormalization, the gauge field propagator is 
represented in the following spectral form (under the substitutions
${\gamma}{\mu}{\equiv}\tilde{\mu}$, $e^2{\gamma}{\equiv}\tilde{e}^2$):
\begin{eqnarray}\label{2.1}
\tilde{D}^{(1)}_{\mu\nu}(p)&=&-i\left(g_{\mu\nu}-\frac{p_{\mu}p_{\nu}}{p^2}\right) 
\gamma\left[\frac{\tilde{Z}_3}{p^2-\tilde{\mu}^2_{\rm ph}+i\epsilon}+
\tilde{\Pi}^{(1)}(p^2)\right] \nonumber \\[2mm]
 &+& \gamma\tilde{\mu}_{\rm ph}\epsilon_{\mu\nu\alpha}
\frac{p^\alpha}{p^2}
\left[\frac{\tilde{Z}_3}{p^2-\tilde{\mu}^2_{\rm ph}+i\epsilon}+
\tilde{\Pi}^{(2)}(p^2)\right].
\end{eqnarray}
The physical mass $\tilde{\mu}_{\rm ph}$ is given by
\begin{equation}\label{2.2}
\tilde{\mu}_{\rm ph} = \tilde{\mu} - \frac{\tilde{e}^2\tilde{\mu}}{8\pi}
\int_{2m}^\infty \frac{1+(4m/a^2)(m-\tilde{\mu})}{a^2-\tilde{\mu}^2} da +
O(\tilde{e}^4).
\end{equation}
The charge renormalization constant $\tilde{Z}_3$ is equal to
\begin{equation}\label{2.3}
\tilde{Z}_3 = 1 - \frac{\tilde{e}^2}{8\pi}
\int_{2m}^\infty da \,
\frac{(1/a^2)(a^2-2m\tilde{\mu})^2+(2m-\tilde{\mu})^2}{(a^2-\tilde{\mu}^2)^2}
+ O(\tilde{e}^4).
\end{equation}
The continuum contributions are 
\addtocounter{equation}{1}
$$ \label{29a} \tilde{\Pi}^{(1)}(p^2) =
\frac{\tilde{e}^2}{8\pi}
\int_{2m}^\infty da\,
\frac{(1/a^2)(a^2-2m\tilde{\mu})^2+(2m-\tilde{\mu})^2}{(p^2-a^2+i\epsilon)
(a^2-\tilde{\mu}^2)^2}
+ O(\tilde{e}^4),\eqno{(27a)} $$
$$ \label{29b} \tilde{\Pi}^{(2)}(p^2) =
\frac{\tilde{e}^2}{4\pi} \left(1-\frac{2m}{\tilde{\mu}}\right)
\int_{2m}^\infty da\,
\frac{a^2-2m\tilde{\mu}}{(p^2-a^2+i\epsilon)(a^2-\tilde{\mu}^2)^2}
+ O(\tilde{e}^4),\eqno{(27b)}$$
The calculation gives:
\begin{eqnarray}
&& Z_3{\equiv}\lim_{\gamma\to\infty}\tilde{Z}_3 = 1, \nonumber \\
&& \label{2.5} \tilde{\mu}_{\rm ph}\equiv\gamma\mu_{\rm ph} =
\gamma\left(\mu-\frac{e^2}{4\pi}\right), \\
&& \label{2.6} {\Pi}^{(1)}(p^2){\equiv}
\lim_{\gamma\to\infty} \gamma\tilde{\Pi}^{(1)}(p^2) =
\frac{e^2}{8\pi\mu^2_{\rm ph}}
\int_{2m}^\infty\frac{da(1+4m^2/a^2)}{a^2-p^2-i\epsilon},
\nonumber \\
&&\lim_{\gamma\to\infty}\gamma\tilde{\Pi}^{(2)}(p^2) = 0.
\end{eqnarray}
Thus
\begin{equation}\label{2.7}
D^{(1)}_{\mu\nu}(p){\equiv}\lim_{\gamma\to\infty}\tilde{D}^{(1)}_{\mu\nu}(p)
=-\frac{1}{\mu_{\rm ph}}
\epsilon_{\mu\nu\alpha}
\frac{p^\alpha}{p^2}-i(g_{\mu\nu}-\frac{p_{\mu}p_{\nu}}{p^2}){\Pi}^{(1)}(p^2).
\end{equation}
The first term coincides with the tree approximation for ${D}^{(0)}_{\mu\nu}$
modulo a finite renormalization of the statistical parameter $\mu_{\rm ph}
=\mu-{e^2}/{4\pi}$. The crucial feature of  Eq.(\ref{2.7})
is the appearance of the parity-even 
term $\sim g_{\mu\nu}-p_{\mu}p_{\nu}/p^2$ in the one-loop 
approximation. This term has no pole in the complex plane of $p^2$.

\subsection{Electron Self-Energy}

After the substitution $\gamma e^2\equiv \tilde{e}^2$,
the spectral form of the fermion
propagator will read (see Eqs.(2.70)--(2.71) of Ref.$\cite{djt}$):
\begin{equation}\label{2.8}
\tilde{S}^{(1)}(p) = i\left[\frac{\tilde{Z}_2}{\hat{p}-\tilde{m}_{\rm ph}}
+\tilde{\Sigma}(p) \right].
\end{equation}
The physical mass, $\tilde{m}_{\rm ph}$, is
\begin{eqnarray} \label{2.9}
&& \tilde{m}_{\rm ph} = m + \frac{\tilde{e}^2}{16\pi}
\int_{-\infty}^\infty da\biggl[
\frac{(\tilde{\mu}+2m)(\tilde{\mu}+2a)}{a^2(a-m)}
\theta(a^2-M^2) \nonumber \\
&&\quad{}+\frac{(a+m+2\tilde{\mu})(a^2-m^2)}{\tilde{\mu}^2a^2}
\theta(M^2-a^2)\theta(a^2-m^2)\biggr] + O(\tilde{e}^4).
\end{eqnarray}
The fermionic renormalization constant, $\tilde{Z}_2$, is given by
\begin{eqnarray} \label{2.10}
&& \tilde{Z}_2 = 1 - \frac{\tilde{e}^2}{16\pi}
\int_{-\infty}^\infty da\biggl[
\frac{(\tilde{\mu}+2m)(\tilde{\mu}+2a)}{a^2(a-m)^2}
\theta(a^2-M^2) \nonumber \\
&&\quad{}+\frac{(a+m+2\tilde{\mu})(a+m)}{\tilde{\mu}^2a^2}
\theta(M^2-a^2)\theta(a^2-m^2)\biggr] + O(\tilde{e}^4),
\end{eqnarray}
where $M=\tilde{\mu}+m$.
The continuum contribution in Eq.(\ref{2.8}) is
\begin{eqnarray} \label{2.11}
 \tilde{\Sigma}(p) &=& \frac{\tilde{e}^2}{16\pi}
\int_{-\infty}^\infty {da\over \hat{p}-a}\biggl[
\frac{(\tilde{\mu}+2m)(\tilde{\mu}+2a)}{a^2(a-m)^2}
\theta(a^2-M^2) \nonumber \\
&+&\frac{(a+m+2\tilde{\mu})(a+m)}{\tilde{\mu}^2a^2}
\theta(M^2-a^2)\theta(a^2-m^2)\biggr] + O(\tilde{e}^4).
\end{eqnarray}
Considering the limit $\gamma\to\infty$ in 
Eqs.(\ref{2.8})--(\ref{2.11}), we get
\begin{eqnarray}
&& \label{2.12} m_{\rm ph}{\equiv}\lim_{\gamma\to\infty} \tilde{m}_{\rm ph}
 = m - \frac{e^2}{2\pi}
\left(\gamma+\frac{m}{3\mu}\right), \nonumber\\[2mm]
&& \label{2.13} Z_2{\equiv}\lim_{\gamma\to\infty} \tilde{Z}_2 = 1 + \frac{e^2}{4\pi}
\frac{5}{3\mu}, \nonumber\\[2mm]
&& \label{2.14} \tilde{\Sigma}_R(p){\equiv}\lim_{\gamma\to\infty}\tilde{\Sigma}(p) 
= \frac{e^2}{4\pi\mu}\frac{\hat{p}}{p^2}
\left(1-\frac{\hat{p}+m_{\rm ph}}{2p}\ln\frac{1+p/m_{\rm ph}}{1-p/m_{\rm ph}}
\right).
\end{eqnarray}
One notices that $\tilde{\Sigma}(p)$ in Eq.(\ref{2.11}) can be represented as
\begin{equation}\label{2.15}
\tilde{\Sigma}(p) = \tilde{\Sigma}_1(p) + \tilde{\Sigma}_2(p),
\end{equation}
where
\addtocounter{equation}{1}
$$\tilde{\Sigma}_1(p) = \frac{\tilde{e}^2}{16\pi}\int_{-\infty}^\infty
\frac{da}{\hat{p}-a}\biggl[\left(\frac{\tilde{\mu}^2}{a^2}+
\frac{4m}{a}\right)\frac{\theta(a^2-M^2)}{(m-a)^2} $$ 
$$\label{2.16a}\quad{}+\frac{1}{\tilde{\mu}^2a^2}(a+m)^2\theta(M^2-a^2)
\theta(a^2-m^2)\biggr], \eqno{(37a)} $$  
$$\tilde{\Sigma}_2(p) = \frac{\tilde{e}^2}{8\pi}\int_{-\infty}^\infty
\frac{da}{\hat{p}-a}\biggl[\frac{(a+m)}{(a-m)^2}\frac{\tilde{\mu}}{a^2}
\theta(a^2-\hat{M}^2) $$  
$$\label{2.16b} \quad{}+\frac{(a+m)}{\tilde{\mu}a^2}\theta(M^2-a^2)
\theta(a^2-m^2)\biggr], \eqno{(37b)}$$ 
where $\tilde{\Sigma}_1(p)$ arises from the exchange of a conventional
transverse vector part of the photon, while $\tilde{\Sigma}_2(p)$ comes from
the axial part of $\tilde{D}^{(0)}_{\mu\nu}$ in the Eq.(\ref{eq5}). 

It is easily shown that
\addtocounter{equation}{1}
$$\label{2.17a}
\lim_{\gamma\to\infty}\tilde{\Sigma}_1(p)=0,
\eqno{(38a)}$$
and thus
$$\label{2.17b}
\lim_{\gamma\to\infty}\tilde{\Sigma}(p)=
\lim_{\gamma\to\infty}\tilde{\Sigma}_2(p) = \tilde{\Sigma}_R(p).
\eqno{(38b)}$$
Therefore, in the limit of the pure CS
spinor electrodynamics  with Pauli-Villars
regularization we get the following result
\begin{equation}\label{2.22}
S^{(1)}(p){\equiv}
\lim_{\gamma\to\infty} \tilde{S}^{(1)}(p)=\frac{iZ_2}{\hat{p}-m_{\rm ph}}+
i\tilde{\Sigma}_R(p).
\end{equation}

Comparing the corresponding results with those in Sect.II, we can see 
that they are the same.

\section{On-shell Vertex Correction}

The one-loop on-shell vertex correction is given by 
\begin{eqnarray}
-i \bar{u}(p')\tilde{\Gamma}_\mu(p', p, m ) u(p)=
\tilde{J}_\mu^a +\tilde{J}_\mu^b +\tilde{J}_\mu^c,
\end{eqnarray}
where
\begin{eqnarray}
\label{2.29a}
\tilde{J}_\mu^a & = & - e^2{\gamma}\int \frac{d^3 q}{(2\pi)^3}
\frac{[-\hat{q}\gamma_\lambda+2(p'+q)_\lambda]\gamma_\mu
[-\gamma^\lambda\hat{q}+2(p+q)^\lambda]}
{(q^2-{\mu}^2{\gamma}^2)\left[(p'+q)^2-m^2
\right]\left[(p+q)^2-m^2\right]}, \\[2mm]
\label{2.29b}
\tilde{J}_\mu^b & = & e^2{\gamma}\int \frac{d^3 q}{(2\pi)^3}
\frac{[-q^2+2(p'+q){\cdot}q]\gamma_\mu
[-q^2 + 2(p+q){\cdot}q]}{q^2(q^2-{\mu}^2{\gamma}^2)\left[(p'+q)^2-m^2
\right])\left[(p+q)^2-m^2\right]}, \\[2mm]
\label{2.29c}
\tilde{J}_\mu^c & = & - e^2{\gamma} \int \frac{d^3 q}{(2\pi)^3}
\frac{i{\mu}{\gamma}\epsilon_{\lambda\sigma\rho}q^\rho[-\hat{q}\gamma^\sigma
+2(p'+q)^\sigma]\gamma_\mu[-\gamma^\lambda\hat{q}+2(p+q)^\lambda]}{q^2
(q^2-{\mu}^2{\gamma}^2)\left[(p'+q)^2-m^2
\right]\left[(p+q)^2-m^2\right]}.
\end{eqnarray}
For derivation of Eqs.(\ref{2.29a})--(\ref{2.29c}), we have used the 
on-shell condition $\hat{p}=\hat{p'}=m$.
The term $\tilde{J}_\mu^b$ is very simple, 
\begin{eqnarray}\label{2.30}
\tilde{J}_\mu^b
={\gamma}e^2\gamma_\mu \int \frac{d^3q}{(2\pi)^3}
\frac{1}{(q^2-{\mu}^2{\gamma}^2)q^2}
=\frac{ie^2}{4\pi{\mu}}\gamma_\mu{\equiv}J_{\mu}^b.
\end{eqnarray}
The term $\tilde{J}_\mu^a$ can be transformed into the following form
\begin{eqnarray}\label{2.31}
\tilde{J}_\mu^a
=-\gamma{e}^2\int \frac{d^3q}{(2\pi)^3}
\frac{\left[q^2\gamma_\mu-2\hat{q}q_\mu+4(p{\cdot}p'+p{\cdot}q+p'{\cdot}q)
\gamma_\mu+4q_\mu m - 4\hat{q}{\cal P}_\mu\right]}
{(q^2-{\mu}^2{\gamma}^2)
\left[(p'+q)^2-m^2\right]
\left[(p+q)^2-m^2\right]},
\end{eqnarray}
where ${\cal P}_\mu \equiv (p'+p)_\mu$. One can not take the limit
${\gamma}{\to}{\infty}$ directly except the term $4p'{\cdot}p$, which vanishes
after the large-$\gamma$ limit. However, using the following decomposition
\begin{eqnarray}
\frac{1}{[(k+p)^2-m^2]}=
\frac{1}{k^2-m^2}-\frac{2k{\cdot}p+p^2}{(k^2-m^2)[(k+p)^2-m^2]}, 
\label{eqde}
\end{eqnarray}  
one can see that all the terms in Eq.(\ref{2.31}) $\sim q$ in 
the numerator vanish when $\gamma\to
\infty$. The first two terms in Eq.(\ref{2.31}) can be transformed into
\begin{eqnarray}
\label{2.32}
\tilde{J}_\mu^a &=& - \gamma {e}^2 \int \frac{d^3 q}{(2\pi)^3}\left\{
\frac{q^2\gamma_\mu-2\hat{q}q_\mu}{(q^2-{\mu}^2{\gamma}^2)(q^2-m^2)^2}\left[
1+\frac{(2p'{\cdot}q+m^2)(2p{\cdot}q+m^2)}{(2p'{\cdot}q+q^2)
(2p{\cdot}q+q^2)}\right.\right.  \nonumber \\[2mm]
&-&\left.\left.\frac{2p'{\cdot}q+m^2}{2p{\cdot}q+q^2}-
\frac{2p'{\cdot}q+m^2}{2p'{\cdot}q+q^2}\right]\right\}.
\end{eqnarray}
Only the first term in Eq.(\ref{2.32}) does not vanish after taking
the limit $\gamma\to\infty$. Thus,
\begin{eqnarray}\label{2.33}
J_\mu^a &\equiv& \lim_{\gamma\to\infty}\tilde{J}_\mu^a =
\lim_{\gamma\to\infty} \gamma {e}^2 \int \frac{d^3 q}{(2\pi)^3}
\frac{2\hat{q}q_\mu-q^2\gamma_\mu}{(q^2-{\mu}^2{\gamma}^2)(q^2-m^2)^2}
\nonumber \\[2mm]
&=& -\lim_{\gamma\to\infty}\frac{{\gamma}{e}^2}{3}\gamma_{\mu} \int
\frac{d^3q}{(2\pi)^3}
\frac{q^2}{(q^2-{\mu}^2{\gamma}^2)(q^2-m^2)^2}  \nonumber \\[2mm]
&=&-\lim_{\gamma\to\infty}\frac{{\gamma}e^2}{3}{\gamma}_{\mu}
\frac{2(\mu\gamma)^3-3(\mu\gamma)^2m+m^3}{8\pi(m^2-{\mu}^2{\gamma}^2)^2}
=-\frac{ie^2}{4\pi\mu}\frac{1}{3}\gamma_\mu.
\end{eqnarray}
As for the third term $\tilde{J}_{\mu}^c$, taking into account 
that in Eq.(\ref{2.29c})
$$ {\epsilon}_{\lambda\sigma\rho}q^\rho\hat{q}{\gamma}^{\sigma}
\gamma_\mu{\gamma}^{\lambda}\hat{q} = -2iq_\mu q^2,$$
and after some algebraic manipulation, we have\footnote{In the numerator of
Eq.(\ref{2.35}) we skip the term $\sim \epsilon_{\lambda\sigma\rho}
q^\rho {p'}^{\sigma} p^{\lambda}$ coming from Eq.(\ref{2.29c}), 
since after integration it will become of the form 
$\epsilon_{\lambda\sigma\rho}p^\rho p^\lambda {p'}^\sigma$ and 
${\epsilon}_{\lambda\sigma\rho}{p'}^{\rho}p^{\lambda}{p'}^{\sigma}$, 
both giving zero.}
\begin{eqnarray}\label{2.35}
\tilde{J}_\mu^c&=& 2e^2{\mu}{\gamma}^2\int \frac{d^3q}{(2\pi)^3}
\frac{[-q_\mu q^2 + (q{\cdot}p') \gamma_\mu \hat{q} + (q{\cdot}p) 
\hat{q}\gamma_\mu+
2m \gamma_\mu q^2-2q^2{\cal P}_\mu]}{q^2(q^2-\tilde{\mu}^2)[(p'+q)^2-m^2]
[(p+q)^2-m^2]} \nonumber \\[2mm]
&=& 2 e^2 \mu{\gamma}^2\int \frac{d^3q}{(2\pi)^3}
\frac{-2q_\mu q^2+(2m\gamma_\mu-2{\cal P}_\mu)q^2+\left[
\hat{q}\gamma_\mu(2p{\cdot}q+q^2)/2+\gamma_\mu\hat{q} (2p'{\cdot}q+q^2)/2
\right]}{q^2(q^2-{\mu}^2{\gamma}^2)[(p'+q)^2-m^2][(p+q)^2-m^2]} 
\nonumber \\[2mm]
&=& 2e^2\mu{\gamma}^2\int \frac{d^3q}{(2\pi)^3} \left\{
\frac{2m\gamma_\mu-2q_\mu-2{\cal P}_\mu}{(q^2-{\mu}^2{\gamma}^2)
(2p'q+q^2)(2pq+q^2)} +\frac{\gamma_\mu\hat{q}}{2q^2(q^2-\mu^2{\gamma}^2)
[(p+q)^2-m^2]} \right.\nonumber \\[2mm]
&+&\left.\frac{\hat{q}\gamma_\mu}{2q^2(q^2-{\mu}^2{\gamma}^2)
[(p'+q)^2-m^2]}\right\}.
\end{eqnarray}
Similar to Eq.(\ref{2.31}), for the terms ${\sim} q$ 
in Eq.(\ref{2.35}) one cannot take 
the large-${\gamma}$ limit directly, 
we still need first to employ the  manipulation (\ref{eqde}). 
Considering the symmetry of integrand, we get that
\begin{eqnarray}
J^c_{\mu}&{\equiv}&\lim_{{\gamma}{\to}{\infty}}\tilde{J}_{\mu}^c\nonumber\\[2mm]
&=&-\frac{e^2}{\mu}{\int}\frac{d^3q}{(2\pi)^3}\left[
\frac{4(m{\gamma}_{\mu}-{\cal P}_{\mu}-k_{\mu})}{(2p{\cdot}q+q^2)
(2p'{\cdot}q+q^2)}+\frac{{\gamma}_{\mu}\hat{q}}{
q^2(2p{\cdot}q+q^2)}+\frac{{\gamma}_{\mu}\hat{q}}{q^2(2p'{\cdot}q+q^2)}
\right].
\end{eqnarray}
The standard Feynman integration gives that
\begin{eqnarray}\label{2.36}
J_\mu^c 
&=& \frac{ie^2}{4\pi\mu}
\left[\gamma_\mu - \frac{2m\gamma_\mu-{\cal P}_\mu}{K}
\ln\frac{1+K/(2m)}{1-K/(2m)}\right]
\nonumber \\[2mm]
&=& \frac{ie^2}{4\pi\mu}
\left[\gamma_\mu - \frac{i\epsilon_{\mu\nu\lambda}
K^\nu\gamma^\lambda}{K}
\ln\frac{1+K/(2m)}{1-K/(2m)}\right],
\end{eqnarray}
where $K_{\mu}{\equiv}p'_{\mu}-p_{\mu}$, $K{\equiv}\sqrt{K^2}$ and 
we have used the three-dimensional analogue of the Gordon identity:
\begin{eqnarray}
{\gamma}_{\mu}=\frac{1}{2m}\left[{\cal P}_{\mu}
+i{\epsilon}_{\mu\nu\lambda}K^{\nu}
{\gamma}^{\lambda}\right].
\label{go}
\end{eqnarray}
Thus at $\gamma\to\infty$, from the eqs.(\ref{2.30}), (\ref{2.33})
and (\ref{2.36}) we get
\begin{eqnarray}\label{2.37}
\lim_{\gamma\to\infty}\left(-i\tilde{\Gamma}_\mu(K)\right)&{\equiv}&
-i\Gamma_\mu(K)=J_\mu^a+J_\mu^b+J_\mu^c,  \nonumber \\[2mm]
{\Gamma}_\mu(K)&=&-\frac{e^2}{4\pi\mu}\left[\frac{5}{3}{\gamma}_{\mu}-
\frac{i\epsilon_{\mu\nu\lambda}K^\nu\gamma^\lambda}{K}
\ln\frac{1+K/(2m)}{1-K/(2m)}\right]\nonumber\\[2mm]
&=&{\gamma}_{\mu}F_1(K^2)+i{\epsilon}_{\mu\nu\lambda}K^{\nu}{\gamma}^{\lambda}
F_2(K^2).
\end{eqnarray}
The vertex renormalization is defined as
\begin{eqnarray}
{\Gamma}_{\mu}(K)={\gamma}_{\mu}(Z_1^{-1}-1)+Z_1^{-1}
{\Gamma}^R_{\mu}(K)
\end{eqnarray} 
and the renormalization condition is as usual 
\begin{eqnarray}
{\Gamma}_{\mu}^R(K)|_{
\hat{p}=\hat{p}'=m,~
K_{\alpha}=p'_{\alpha}-p_{\alpha}=0}=0.
\end{eqnarray}
Then we get  the vertex renormalization constant
\begin{eqnarray}
Z_{1}^{-1}{\gamma}_{\mu}&=&{\gamma}_{\mu}+{\gamma}_{\mu}F_1(0),
\nonumber\\[2mm]
{Z_1}^{-1}&=&1-\frac{e^2}{4\pi\mu}\frac{5}{3},
\end{eqnarray}
and the one-loop radiative correction to the vertex as
\begin{eqnarray}
{\Gamma}_{\mu}^R(K)
&=&-{\gamma}_{\mu}+Z_1({\gamma}_{\mu}+{\Gamma}_{\mu})
\nonumber\\[2mm]
&=&\frac{ie^2}{4{\pi}\mu}{\epsilon}_{\mu\nu\lambda}K^{\nu}{\gamma}^{\lambda}
\frac{1}{K}\ln\frac{1+K/(2m)}{1-K/(2m)}.
\end{eqnarray}
 From Eq.(\ref{eq25}) we have
\begin{eqnarray}\label{2.39}
{Z_1}=1+\frac{e^2}{4\pi\mu}\frac{5}{3}=Z_2,
\end{eqnarray}
which is just the consequence of Ward identity 
\begin{eqnarray}
K^{\mu}{\Gamma}_{\mu}(K)={\Sigma}(p')-{\Sigma}(p).
\end{eqnarray}
It is remarkable that ${\Gamma}_{\mu}^R(K^2=0)$ does not vanish, i.e.
\begin{eqnarray}
{\Gamma}_{\mu}^R(0)=\frac{ie^2}{4\pi}
\frac{1}{{\mu}m}{\epsilon}_{\mu\nu\lambda}
K^{\nu}{\gamma}^{\lambda}
=i\frac{\alpha}{{\mu}m}{\epsilon}_{\mu\nu\lambda}K^{\nu}{\gamma}^{\lambda},
\end{eqnarray}
which gives the three-dimensional analogue  
of Schwinger's result for the anomalous magnetic moment
of the electron. In a slowly varying (in both space and time) 
external electricmagnetic field, it will
lead to a new interaction  Hamiltonian\footnote{The self-energy insertion in the external
line can be disregarded since the electrons are on mass-shell.}:
\begin{eqnarray}
{\Delta}{\cal H}&=&-\frac{\alpha}{m{\mu}}{\epsilon}^{\mu\nu\lambda}
\bar{\psi}(x){\gamma}_{\lambda}{\psi}(x)
{\partial}_{\nu}A_{\mu}=-\frac{\alpha}{2m{\mu}}{\epsilon}^{\mu\nu\lambda}
\bar{\psi}(x){\gamma}_{\lambda}{\psi}(x)F_{\mu\nu}\nonumber\\[2mm]
&=&-\frac{\alpha}{2m{\mu}}\bar{\psi}(x)
{\sigma}^{\mu\nu}{\psi}(x)F_{\mu\nu},
\end{eqnarray}
where we have used that 
\begin{eqnarray}
{\epsilon}_{\mu\nu\lambda}{\gamma}^{\lambda}
=\frac{i}{2}[{\gamma}_{\mu},{\gamma}_{\nu}]{\equiv}{\sigma}_{\mu\nu}.
\end{eqnarray}
Thus this term leads to the anomalous magnetic 
moment of the electron$\cite{iz}$, which is consistent with the result
in Ref.$\cite{ks}$.
It is very interesting that this term exists in scalar case too
$\cite{ccf}$.

\section{Pure Chern-Simons Electrodynamics}

Now we consider the case of pure CS electrodynamics, i.e.
put ${\gamma}{\to}{\infty}$ at the tree level. The vacuum polarization 
tensor and $D^{(1)}_{\mu\nu}(p)$ will be the same since this does not change 
the electron loop.
However, the electron self-energy and the vertex correction will be different
since the gauge field propagator is replaced by Eq.(\ref{eq6}).

We first  consider the electron self-energy  
\begin{eqnarray}
-i{\Sigma}^{\rm pure}(p)&=&
-\frac{2e^2}{\mu}{\int}\frac{d^nq}{(2\pi)^n}\frac{q^2+(\hat{p}-m)\hat{q}}
{q^2[(q+p)^2-m^2]}\nonumber\\[2mm]
&=&\frac{ie^2}{4\pi\mu}\left\{2m-(\hat{p}-m)\frac{\hat{p}}{m}\left[
\frac{m^2}{p^2}+\frac{m^3}{2p^3}\left(1-\frac{p^2}{m^2}\right)
\ln\frac{1-p/m}{1+p/m}\right]\right\}.
\end{eqnarray}
Similar discussions as the ones used in getting Eq.(\ref{eq25}) give that
\begin{eqnarray}
m_{\rm ph}^{\rm pure}&=&m(1+\frac{e^2}{2\pi}\frac{1}{\mu}), 
~~Z_2^{\rm pure}=1+\frac{e^2}{4\pi}\frac{1}{\mu},\nonumber\\[2mm]
{\Sigma}^{\rm pure}_R(p)&=&
-\frac{e^2}{4\pi\mu}(\hat{p}-m_{\rm ph})\left\{\frac{\hat{p}}{m_{\rm ph}}\left[
\frac{m_{\rm ph}^2}{p^2}
+\frac{m_{\rm ph}^3}{2p^3}\left(1-\frac{p^2}{m_{\rm ph}^2}\right)
\ln\frac{1-p/m_{\rm ph}}{1+p/m_{\rm ph}}\right]-1\right\},\nonumber\\[2mm]
\tilde{\Sigma}^{\rm pure}_R(p)
&=&\frac{e^2}{4\pi}\frac{1}{\mu}\frac{\hat{p}}{p^2}\left[
1+\frac{\hat{p}+m_{\rm ph}}{2p^3}\ln\frac{1-p/m_{\rm ph}}{1-p/m_{\rm ph}}\right].
\end{eqnarray}

Using the techniques stated above, the on-shell vertex correction is given
as follows
\begin{eqnarray}
&-&i\bar{u}(p'){\Gamma}^{\rm pure}_{\mu}(p',p,m)u(p){\equiv}
-i{\Gamma}^{\rm pure}_{\mu}(K)
=\frac{ie^2}{\mu}{\int}\frac{d^nq}{(2\pi)^n}
\frac{{\gamma}_{\rho}(\hat{q}+\hat{p}'+m){\gamma}_{\mu}(\hat{q}+
\hat{p}+m){\gamma}_{\nu}{\epsilon}^{\nu\rho\lambda}q_{\lambda}}{
q^2[(q+p')^2-m^2][(q+p)^2-m^2]}\nonumber\\[2mm]
&=&-\frac{2e^2}{\mu}{\int}\frac{d^nq}{(2\pi)^n}\left[
\frac{{\gamma}_{\mu}\hat{q}}{2q^2(q^2+2p{\cdot}q)}+
\frac{\hat{q}{\gamma}_{\mu}}{2q^2(q^2+2p'{\cdot}q)}+
\frac{2m{\gamma}_{\mu}-2{\cal P}_{\mu}-2q_{\mu}}{(q^2+2p'{\cdot}q)(
q^2+2p{\cdot}q)}\right]\nonumber\\[2mm]
&=&\frac{ie^2}{4\pi\mu}\left[{\gamma}_{\mu}-i{\epsilon}_{\mu\nu\lambda}
K^{\nu}{\gamma}^{\lambda}\frac{1}{K}\ln\frac{1+K/(2m)}{1-K/(2m)}\right],
\end{eqnarray}
where the three-dimensional Gordon identity (\ref{go}) has been used.
Correspondingly, the vertex renormalization constant is
\begin{eqnarray}
Z_1^{\rm pure}=1+\frac{e^2}{4\pi}{\gamma}_{\mu},
\end{eqnarray}
and we  still have 
\begin{eqnarray}
Z_1^{\rm pure}=Z_2^{\rm pure}.
\end{eqnarray}
In particular, we still obtain the same anomalous magnetic moment term.  

\section{Conclusion and Discussion}

We have made a detailed study of the quantum correction
to CS spinor electrodynamics.  We give complete analytical
results for one-loop quantum corrections such as polarization tensor,
electron self-energy and specially for the on-shell vertex. We find the three
dimensional analogue of the Schwinger anomalous magnetic
term, despite it is in the second order, this may lead to nontrivial
planar dynamics since it can provide new interaction between
charged particles. We compare the different procedure of
taking the limit ${\gamma}{\rightarrow}{\infty}$ and verify explicitly
that in both cases the Ward identity is satisfied, and hence that the physical
quantities are independent of the order of taking large-$\gamma$ limit. 

In addition, in both cases, the results are finite and
the ${\beta}$-function vanishes identically.
If we take into account the higher order perturbative corrections, 
according to BPHZ renormalization procedure, 
we believe that the results 
are still finite since the one-loop renormalization constants
are all finite and all propagators and vertex part in the asymptotic 
region will be the same as those in the free case after renormalization. 

\acknowledgments
We thank Dr. R. Ray for useful correspondence.
The  financial support of the Academy of Finland is greatly
acknowledged. W.F.C. thanks the World 
Laboratory, Switzerland for financial support and V.Ya.F. thanks
for the financial support by RFBR Grant No. 96-01-00105.

\end{document}